\documentclass[%
aip,
 amsmath,amssymb,
reprint,%
]{revtex4-1}

\usepackage{multirow}
\usepackage{graphicx}
\usepackage{bm}% bold math
\usepackage{hyperref}% add hypertext capabilities
\usepackage[version=3]{mhchem} 
\usepackage{chemarr}

     \makeatletter
\newcommand\xleftrightarrow[2][]{%
  \ext@arrow 9999{\longleftrightarrowfill@}{#1}{#2}}
\newcommand\longleftrightarrowfill@{%
  \arrowfill@\leftarrow\relbar\rightarrow}
\makeatother

\usepackage{subfigure, textcomp, amssymb, mathrsfs}

\usepackage{epstopdf}

\usepackage{mathrsfs}

\begin{document}

\title{Tikhonov regularization for the deconvolution of  capacitance from voltage-charge response of electrochemical capacitors}% with nonlinear charge-voltage characteristics}  

\author{Anis Allagui}
\email{aallagui@sharjah.ac.ae}
\affiliation{Dept. of Sustainable and Renewable Energy Engineering, University of Sharjah, PO Box 27272, Sharjah, United Arab Emirates}
\altaffiliation[Also at ]{Center for Advanced Materials Research, Research Institute of Sciences and Engineering, University of Sharjah, PO Box 27272, Sharjah, United Arab Emirates}
\altaffiliation[and ]{Dept. of Mechanical and Materials Engineering, Florida International University, Miami, FL33174, United States}

\author{Ahmed S. Elwakil}
\affiliation{Dept. of Electrical and Computer Engineering, University of Sharjah, PO Box 27272, Sharjah, United Arab Emirates}
\altaffiliation[Also at ]{Nanoelectronics Integrated Systems Center (NISC), Nile University, Cairo, Egypt}
\altaffiliation[and ]{Dept. of Electrical and Computer Engineering, University of Calgary, Calgary, Canada}
%

%\author{Mohammed E. Fouda}
%\affiliation{Electrical Engineering and Computer Science Dept.,  University of California-Irvine, CA, United States}
%\altaffiliation[Also at ]{Engineering Mathematics and Physics Dept., Faculty of Engineering, Cairo University, Giza, Egypt}

%
\begin{abstract}

The capacitance of capacitive energy storage devices can not be directly measured, but can be estimated from the input and output signals expressed in the time or frequency domains. 
Here the time-domain voltage-charge relationship in non-ideal electrochemical capacitors is treated as an ill-conditioned convolution integral equation where the unknown capacitance kernel function is to be found. This comes from assuming \emph{a priori} that in the frequency domain the charge is equal to the product of capacitance by voltage. The computation of a stable solution to this problem particularly when dealing with experimental data is highly sensitive to noise as it  may lead to an oscillating output even in the presence of small errors in the measurements. In this work, the problem is treated using Tikhonov's regularization method, where a degree of damping is added to each singular value decomposition (SVD) component of the solution, thus effectively filtering out the components corresponding to the small singular values.

%\noindent Keywords : Tikhonov regularization; Electrochemical capacitors;  Constant-phase element 

\end{abstract}

\maketitle

\section{Introduction}

Electrochemical capacitors, also known as electric double-layer capacitors or supercapacitors, can store large amounts of electrical energy in relatively small volumes, and discharge it back in  short times \cite{simon2014batteries}.  
They do so very efficiently with low charge-discharge hysteresis, with extremely long cyclic lifetime, and with excellent safety features and cost effectiveness, which renders them the option of choice for applications where bursts of power are needed \cite{shao2018design}. 
The large energy storage capability of these devices is based on the  electrostatic double layer formed at the interface between high-surface area  porous electrodes in contact with a liquid electrolyte \cite{yang2022understanding}. Specifically, the storage mechanism in response to an external excitation involves ionic transports in opposite directions (physical adsorption of counterions and repulsion of coions) which can face different resistive paths depending on the local structure of the electrodes (porosity, surface roughness, atomic-scale inhomogeneities \cite{KERNER2000207}), the dynamics of the eletric double-layer and the formulation of the electrolytic bath \cite{foedlc}. As a result, a frequency dispersion of capacitance, following a power-law decay of the form $C(\omega) \propto (i\omega)^{\alpha-1}$ with $0<\alpha<1$, is commonly observed in such devices and in other systems involving  electrified electrodes/electrolyte interfaces. 

Now if the capacitance is a frequency-dependent function in the frequency domain, it is natural that   in the time domain it must be treated as a time-dependent function. 
We recall that a frequency-domain representation of a signal in terms of magnitude and phase as a function of frequency is an indirect way of looking at time-domain data using time-to-frequency transformations (e.g.  Fourier or Laplace transforms).   
In the wide literature we see that the same capacitive system is characterized   on the one hand by a constant phase element, and thus a capacitive function in the frequency domain, and on the other hand by a constant capacitance from time-domain measurements (see \cite{acs2} and the references therein). These inconsistencies have been raised and discussed in some of our recent contributions on the topic \cite{acs2, allagui2021inverse, ieeeted, allagui2022time}, which led to renewed interest in  establishing proper characterization methods for capacitive devices \cite{jeltsema2023further, ortigueira2023new}. It is needless to mention that erroneous calculation of capacitance has direct implications on the computation of energy and power performance of electrochemical capacitors needed for their proper integration in larger circuit systems. 

In this work we analyze the experimental time-domain charge-voltage response of a non-ideal electric double-layer capacitor that shows a fractional impedance behavior in the frequency domain of the constant phase element type. The charge-voltage in the frequency domain are assumed to be related via a multiplicative capacitive function, which is in line with the definition of impedance function for linear systems. Thus,  the corresponding input-output time-domain problem is a convolution of a capacitive function with the applied voltage resulting in the measured charge, from which we extract by using Tikhonov regularization method the characteristic function of the device. This is a continuation of our previous work on the same problem \cite{allagui2021inverse}, wherein  we focused mostly on the transition between time- and frequency-domain definitions of capacitance functions, but did not go far enough with the proper analysis of real experimental data that are unavoidably  corrupted with noise.

\begin{figure*}[!t]
\begin{center}
\includegraphics[width=0.9\textwidth]{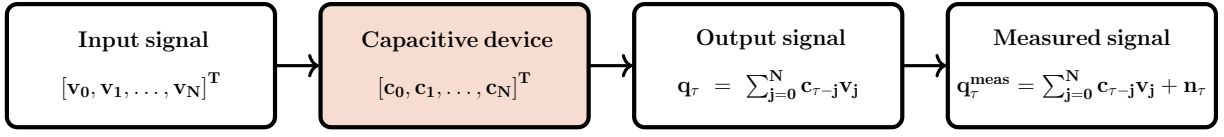}
\caption{Linear input-output convolution model of a capacitive system and the effect of noise in the measured signal}
\label{fig1}
\end{center}
\end{figure*}
\section{Methods}
 
\subsection{Theory}

Electrochemical capacitors are assumed to be linear time-invariant (LTI), causal and stable systems that map an input voltage signal $\mathbf{v[t]}$ to an output  charge $\mathbf{q[t]}$. Both input and output are real time-domain signals. We rewrite the input voltage signal as a sequence of discrete values such that $\mathbf{v} = \left[\bf v_0,  v_1, \hdots,  v_N \right]^{\text{T}}$ (T for transpose) of finite length, and the same for the charge 
$\mathbf{q} = \left[\bf q_0,  q_1,  \hdots , q_N \right]^{\text{T}} = \left[f(v_0) , f(v_1) , \hdots , f(v_N) \right]^{\text{T}}$. Both voltage and charge vectors are assigned respectively to the equidistant time instances $0^+, t_1, \ldots, t_N$ ($t_i=i \Delta t$, we assume $\Delta t=1$). 
 When a voltage $\bf v_0$ is applied on an capacitive device at $t=0^+$, the  corresponding measured charge  is $\bf q_0=c_{0} v_0  $, where $\bf c_0$ is the value of the system impulse response at that instant.  The subsequent amount of charge $\bf q_1$ accumulated on the device after a lapse of time $\Delta t$  becomes the sum  of $\bf c_{0} v_1  $ and $ \bf c_{1} v_0 $ by considering the superposition principle in causal LTI systems. This can be understood by realizing that the input voltage  at time  $\Delta t$ has now the value of $\bf v_1$ which, with the impulse response of the device $\bf c_0$ results in a certain amount of charge $\bf c_{0} v_1  $, but the preceding value of voltage $\bf v_0$ adds also  another amount of charge $\bf c_{1} v_0 $  that depends on the current response of the device $\bf c_1$ at the instant $\Delta t$. Repeating this procedure for all discrete times $t_i=i \Delta t$ until $N\Delta t$ (total time duration of the applied signal) is  the discrete convolution sum of the input voltage by the systems impulse response which results in the output charge as:   
\begin{equation}
\bf q_\tau = (c * v)_\tau = \sum\limits_{i=0}^{N} c_{{\tau - i}} v_i  
\label{eq:qn} 
\end{equation}
A schematic illustration of this process is depicted in 
Fig.\;\ref{fig1}. Eq.\;\ref{eq:qn} can be expressed as a matrix multiplication representing a system of linear equations describing the problem as:
\begin{equation}
\bf q = v_m\, c
\label{eq:q}
\end{equation}
Here $\bf v_m$ is a $(N+1)\times (N+1)$ lower triangular matrix of the Toeplitz matrix, whose columns are shifted versions of the input signal vector $\bf v$ ($\bf v_{i,j}=v_{i-j}$) \cite{gray2006toeplitz}, i.e.
\begin{equation}
\bf{v_m}
=
\begin{bmatrix}
    \bf v_{0} & 0 & \dots   & 0 \\
    \bf v_{1} & \bf v_{0} & \dots   & 0 \\
    \vdots & \vdots  & \ddots  &\vdots \\
    \bf v_{N} & \bf v_{N-1}  & \dots &\bf v_{0}
\end{bmatrix}
\label{matrixv}
\end{equation}
 and the vector $\mathbf{c} = \left[\bf c_0,  c_1, \hdots,  c_N \right]^{\text{T}}$ represents the unknown  model parameter values (capacitance function). 
 
In continuous-time form, this forward problem   is the convolution integral of the input signal with the transfer function of the system, $c(t)$, which is also referred to as the kernel of the convolution operation, such that:
\begin{equation}
q(t) = (c \circledast v) (t) 
:= \int_{0}^t c(t-\tau) v(\tau) d\tau 
\end{equation}
 The function $c(t)$ can be viewed as the system-level, macroscopic capacitance of the device that takes into account (in a lumped form) all microscopic events of charge transport and storage taking place inside the device in response to an applied voltage.

The inverse problem of reconstructing the system response $\mathbf{c} = \left[\bf c_0,  c_1, \hdots,  c_N \right]^{\text{T}}$ from the convolution operation when the input  and output signals are known as an ill-conditioned deconvolution problem for system identification \cite{mueller2012linear}. Solutions to well-posed problems have the properties of existence, uniqueness and stability of the solution (Hadamard criteria), but this is not necessarily the case.    
When the possible sources of input signal distortion have a linear effect on the output signal that can be lumped into the identification of $
\bf c$,  one can obtain from Eq.\;\ref{eq:q} a naive solution as:
\begin{equation}
\bf c = v_m^{-1} \,q
\end{equation}
if the matrix $\bf v_m$ is invertible. If some rows of $\bf v_m$ are close to being linear combinations of each other, the matrix is nearly singular and inversion cannot be done directly.  
Furthermore, when   noise in the input and output  are uncorrelated,  the  problem given by Eq\;\ref{eq:qn} takes instead the form:
\begin{equation}
\bf q_\tau^{\text{meas}} =   \sum\limits_{i=0}^{N} c_{{\tau - i}} v_i  + {n_\tau}
 \end{equation}
 or in matrix notation as:
 \begin{equation}
\bf q^{\text{meas}} = v_m\, c + n = q+n
\label{eq:qm}
\end{equation}
to describe the measured data (Fig.\;\ref{fig1}). 
 The vector $\bf n$ models errors coming from all sources of noise, and can be considered as a random variable with certain statistics: $E(\mathbf{n})=\mathbf{0}$ and $E(\mathbf{n} \,\mathbf{n}^{\text{T}})=\mathbf{S^2}$, where $E(\cdot)$ is the expectation operator and $\mathbf{S^2}$ is the positive definite variance matrix. When the measurement errors are statistically uncorrelated, then the variance matrix is diagonal, i.e. $\mathbf{S^2}=\mathbf{\text{diag}}(s_1^2, s_2^2,\ldots,s_N^2)$ where $s_1^2, s_2^2,\ldots,s_N^2$ are the standard deviations of the errors, otherwise  when the errors are correlated, the problem can still be transformed to have a diagonal variance matrix by premultiplying Eq.\;\ref{eq:qm} by the inverse of the lower triangular Cholesky factor of $\mathbf{S^2}$ \cite{resPeriodograms}.
  Even small perturbations in the signals can result in amplified arbitrary perturbations in the  solution, and in general at least one of the conditions of existence, uniqueness and stability fails for the obtained solution.   
  Therefore, it is usually required to apply some form of a computational regularization method to dampen out such instabilities and arrive to a meaningful result.

We note that our problem is looked at from  the time-domain perspective, but in the frequency domain, it turns out to be more advantageous from a computational point of view. Let $ Q$, $ C$ and $ V$ be the discrete Fourier transforms (DFT) of $\bf q$, $\bf c$ and $\bf v$ respectively, then the discrete time series can be represented by the inverse  DFTs (iDFT) as:
\begin{eqnarray}
\mathbf{v_j} &=& \frac{1}{N} \sum\limits_{m=0}^{N-1} {V_m} \exp \left[ \frac{ 2 i \pi  j m}{N} \right]\\
\mathbf{c_{\tau-j}} &=& \frac{1}{N} \sum\limits_{n=0}^{N-1} {C_n} \exp \left[ \frac{2i\pi (\tau-j) n}{N} \right]\\
\mathbf{q_{\tau}} &=& \frac{1}{N} \sum\limits_{l=0}^{N-1} {C_l} \exp \left[ \frac{ 2 i \pi  \tau l}{N} \right]
\end{eqnarray}
respectively. After a few algebraic manipulations, the original convolution sum in the time domain given in Eq.\;\ref{eq:qn} becomes a point-by-point multiplication  in the frequency domain of the DFTs of the signals, i.e.: 
\begin{equation}
  Q = C V
\end{equation}
The spectrum or DFT of the solution to our problem is thus:
\begin{equation}
C = \frac{Q}{V}
\end{equation} 
which requires applying the iDFT to revert back to the time-domain solution \cite{allagui2022time, ieeeted, allagui2021inverse, acs2}. The DFT and the iDFT of a signal can  be computed efficiently by means of the fast Fourier transform (FFT) algorithm in $O(N \ln N)$ operations \cite{hansen2002deconvolution}.
However,   frequency-domain data are not always available, and usually one has  to work with time-domain data. Secondly, the solution can be        unstable owing to the  strong response of high Fourier harmonics to arbitrary small distortion of data. 

One way to recover $\bf c$ from Eq.\;\ref{eq:qm}, from which we will drop all subscripts and superscripts for ease of notation,
 is to minimize the squares of the errors, i.e.:
 \begin{equation}
r^2_{\min}(\mathbf{c}) = \min_{\mathbf{c}}\{ (\bf q-v\, c)^{\text{T}} (\bf q-v\, c)\}
\end{equation}
 The linear regression estimate  \cite{aster2018parameter}:
 \begin{equation}
\mathbf{\tilde{c}} = \bf (v^{\text{T}} v)^{-1} v^{\text{T}} q
\end{equation}
is the best linear unbiased estimate of the true value of $\mathbf{c}$ \cite{resPeriodograms}. 
However, if the product $\bf (v^{\text{T}} v)$ is close to singular, the problem has
more solutions. The optimal solution giving the least-squared error is found using the singular value decomposition (SVD) method, which factorizes a matrix $\bf v$ (can be rectangular or square matrix) as:
 \begin{equation}
\mathbf{v} = \mathbf{U\Sigma V^{\text{T}}} = \sum\limits_{i=1}^N \mathbf{u}_i\, \sigma_i\, \mathbf{v}_i^T
\end{equation}
Here $\bf U$ and $\bf V^{\text{T}}$ are orthogonal matrices (i.e. $\bf U^{\text{T}} U =V^{\text{T}} V = I$) with their columns ($\mathbf{u}_i$ and $\mathbf{v}_i$ are the $i^{\textit{th}}$ columns of $\mathbf{U}$ and $\mathbf{V}$, respectively) being the left and right singular vectors of $\bf v$, respectively. The matrix $\bf \Sigma$ in between is a diagonal (scaling) matrix whose diagonal elements $\sigma_i$ are the (nonnegative) singular values of $\bf v$ ordered in a decreasing order in the diagonal as $\sigma_1 \geqslant \sigma_2 \geqslant \ldots  \geqslant \sigma_N \geqslant 0$ (note that some of the singular values may be zero). It can be shown that the  number of nonzero singular values is equal to the rank of $\mathbf{v}$,  and that the condition number of $\mathbf{v}$ is $\text{cond}(\mathbf{v}) = \sigma_1/\sigma_N$ \cite{hansen2002deconvolution}. The naive solution to the problem is given by \cite{hansen2010discrete}:
\begin{equation}
\mathbf{c}_n= {\mathbf{V  {\Sigma}^{-1} U^{\text{T}} q}} =\sum_{i=1}^N \frac{\mathbf{u}_i^{\text{T}}\mathbf{q}}{\sigma_i} \mathbf{v}_i
\end{equation}

Otherwise, if noise is involved in the measurements, one can compute an estimate of the solution by simply truncating the sum at some  $k<N$ such that:
\begin{equation}
\mathbf{\tilde{c}}_k = \mathbf{V \tilde{\Sigma}^{-1} U^{\text{T}} q }= \sum_{i=1}^k \frac{\mathbf{u}_i^{\text{T}}\mathbf{q}}{\sigma_i} \mathbf{v}_i
\label{eq15}
\end{equation}
where $\bf \tilde{\Sigma}^{-1}$ is the inverse of $\bf {\Sigma}$ with some elements of $\bf \tilde{\Sigma}^{-1}$  set to zero if the singular value is below a certain threshold, i.e. $\mathbf{\tilde{\Sigma}}=\text{diag}(\sigma_1,\sigma_2,\ldots,\sigma_k,0,\ldots,0)$.  
This means that some  equations corrupted by numerical instabilities  will be  discarded from the overall  system of equations describing the device under test. This transforms the ill-conditioned problem to a well-conditioned one, but rank-deficient. As a result one  should obtain a more robust and less unstable estimate of the solution.  The threshold or the filtering method to be applied, which helps with the regularization of the solution such that components with high frequencies are excluded, is usually related to the quality of the data and the extent of the noise level. This is the so-called truncated SVD (TSVD)   technique, which is probably the simplest direct regularization method that can be applied.

\begin{figure*}[!t]
\begin{center}
\subfigure[]
{\includegraphics[height=1.5in]{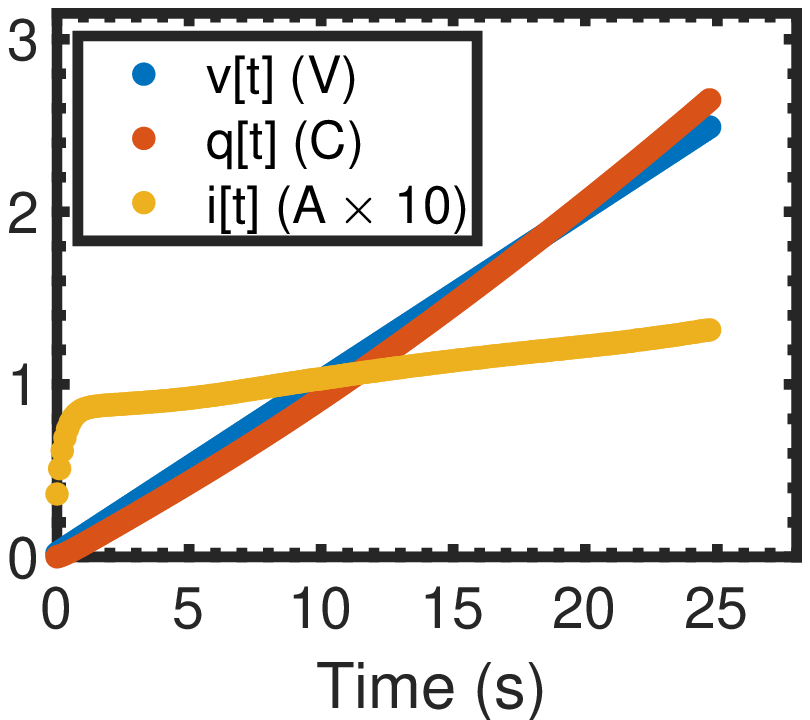}}
\subfigure[]
{\includegraphics[height=1.5in]{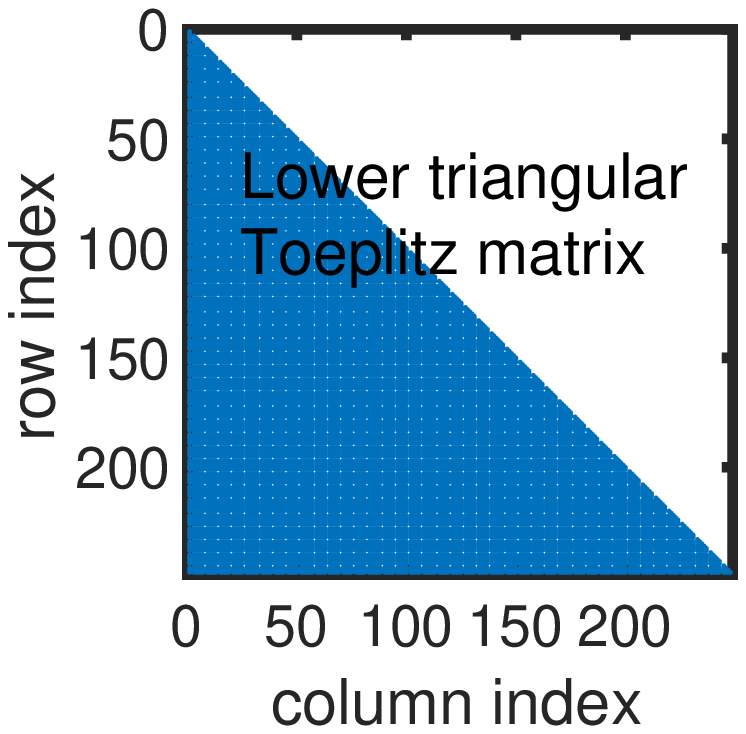}}
\subfigure[]
{\includegraphics[height=1.5in]{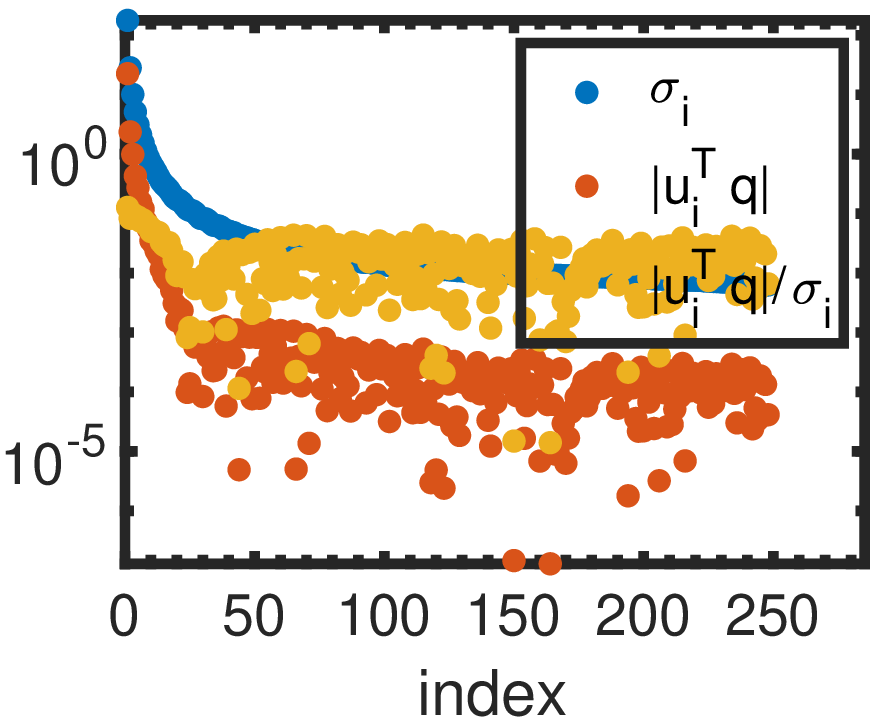}}
\subfigure[]
{\includegraphics[height=1.5in]{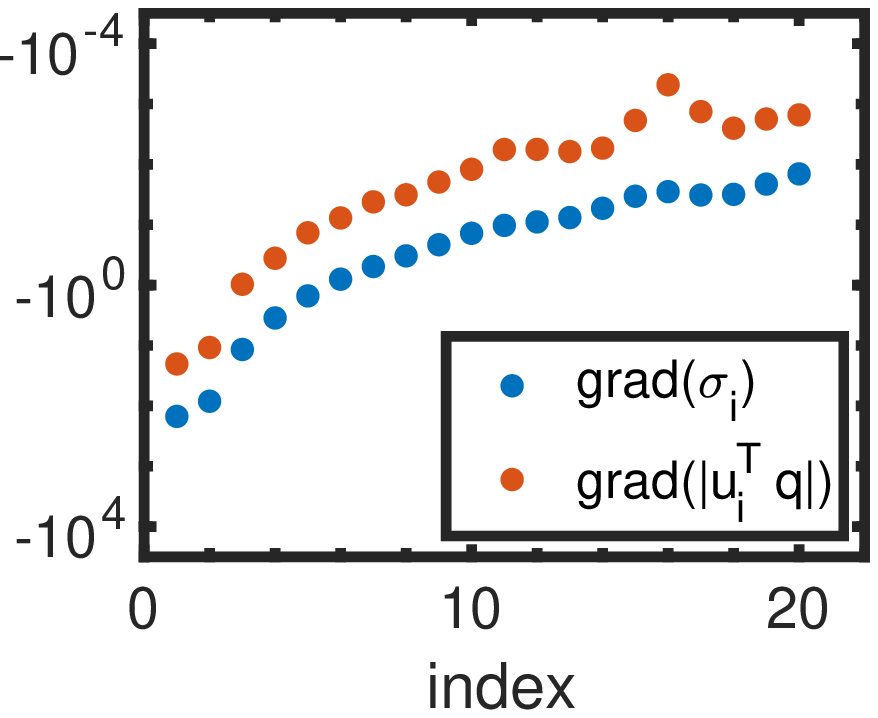}} \\
\subfigure[]
{\includegraphics[height=1.7in]{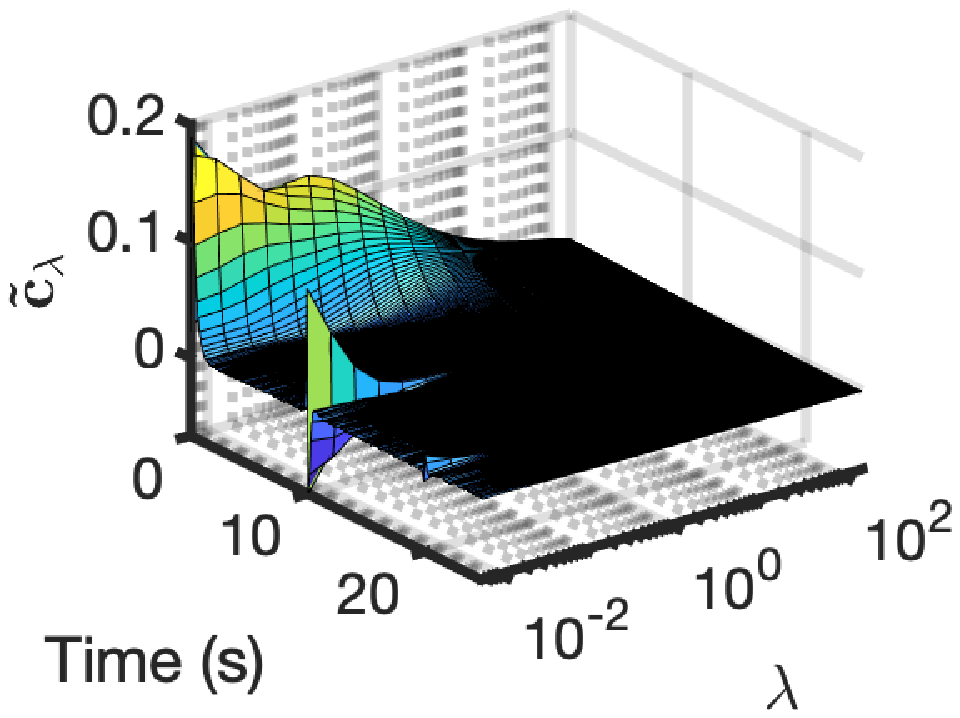}}
\subfigure[]
{\includegraphics[height=1.7in]{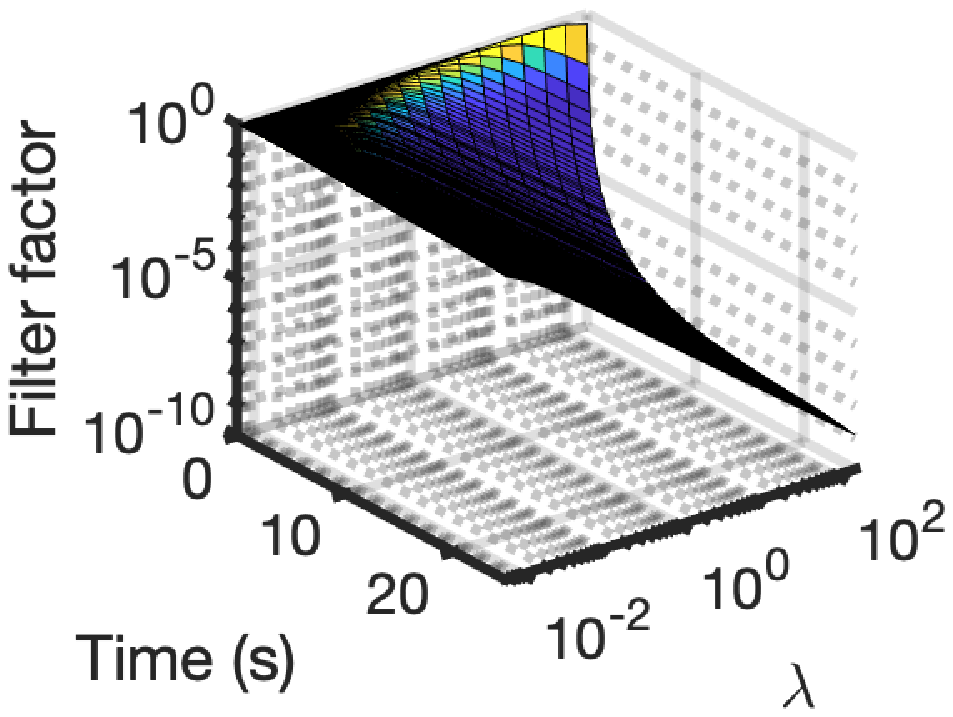}}
\subfigure[]
{\includegraphics[height=1.5in]{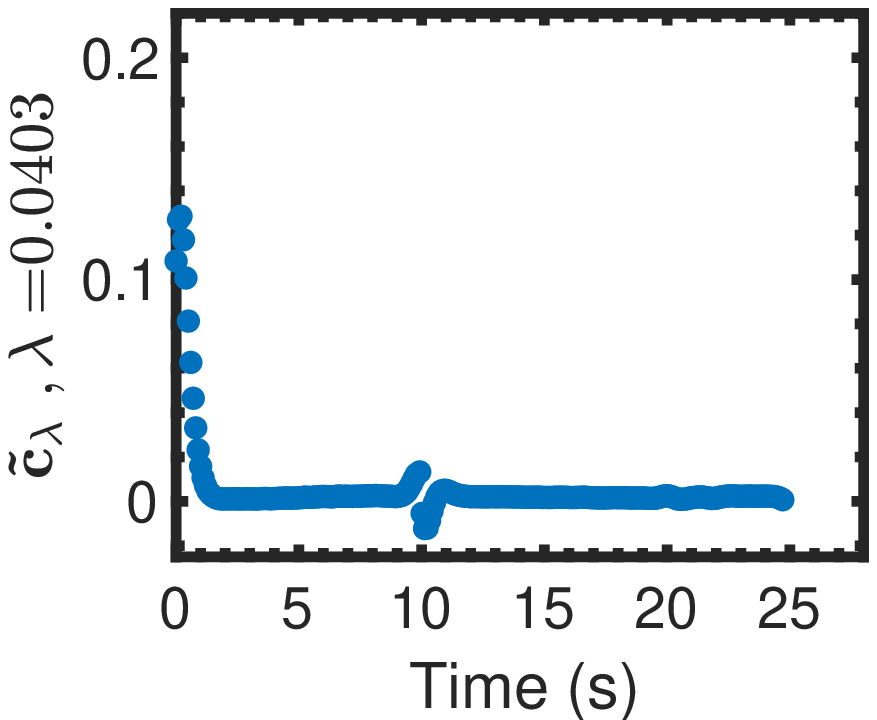}}
\caption{(a) Input voltage and accumulated charge (derived from the current signal) time series on the Samxon EDLC, 
(b) sparsity pattern of the 248$\times$248 lower triangular Toeplitz matrix $\mathbf{v}$  constructed from the voltage time series data in (a), 
(c) Picard plots of the singular values $\sigma_i$ and the SVD-components $|u_i^{\text{T}}\mathbf{q}|$ and $|u_i^{\text{T}}\mathbf{q}/\sigma_i|$ for the matrix $\mathbf{v}$ as a function of index $i$ ($1\leqslant i\leqslant 248$), 
(d) plot of gradients of $\sigma_i$ and $|u_i^{\text{T}}\mathbf{q}|$ for $1\leqslant i\leqslant 20$
(e) Tikhnov regularized capacitance function  $\mathbf{\tilde{c}}_{\lambda}$ for different values of $\lambda$
and (f) their associated filter factors, $f_i= {\sigma_i^2}/{(\sigma_i^2+\lambda^2)}$ in Eq.\;\ref{eq:clambda}. 
(g) Plot of Tikhnov-regularized capacitance function  $\mathbf{\tilde{c}}_{\lambda}$ for $\lambda=0.0403$}
\label{qvt}
\end{center}
\end{figure*}

An alternative way to just setting the small singular values of $\mathbf{v}$ to zero is to instead increase them by small increments so that they lead to a smaller contribution to the solution estimate. This is the Tikhonov regularization method which gives an estimate to the solution as \cite{golub1999tikhonov,hansen2002deconvolution}:
\begin{equation}
\mathbf{\tilde{c}}_{\lambda} = (\mathbf{v}^{\text{T}} \mathbf{v} + \lambda^2 \mathbf{I})^{-1} \mathbf{v}^{\text{T}} \mathbf{q} = \sum\limits_{i=1}^N \left(\frac{\sigma_i^2}{\sigma_i^2+\lambda^2} \right)\frac{\mathbf{u}_i^{\text{T}}\mathbf{q}}{\sigma_i} \mathbf{v}_i
\label{eq:clambda}
\end{equation}
being the result of  the minimization problem:
\begin{equation}
\min_{\mathbf{c}} \{|| \mathbf{q-vc}||^2+\lambda^2 ||\mathbf{c}||^2\}
\end{equation}
Here $\lambda$ is a parameter  to  tradeoff between closeness to the true solution and distance from noise.  
It is clear   that as $\lambda \to 0$, the estimate $\mathbf{\tilde{c}}_{\lambda}$ approaches the naive solution to the problem, whereas as $\lambda \to \infty$, $\mathbf{\tilde{c}}_{\lambda} \to 0$. In   practical applications, $\lambda$ is bounded by the minimum and maximum of the singular values, i.e. $\sigma_N$ and $\sigma_1$. The Tikhonov solutions resemble the TSVD solutions when the truncation parameter $k$ and the regularization parameter $\lambda$ are chosen such that $\sigma_k \approx  \lambda$ \cite{hansen2010discrete}.  We will apply the Tikhonov regularization method in what follows.

\subsection{Experimental}

A  Samxon EDLC (DRL series, part No. DRL105S0TF12RR, rated 2.7\,V, 1.0\,F) is taken as a test device. The electrical measurements are carried out on a Biologic VSP-300 potentiostat equipped with an impedance spectroscopy (EIS) module. The instrument's control voltage is $\pm$10\,V with a  resolution of 1 $\mu$V on 60\,mV range, and its current ranges are 500\,mA to 10\,nA with a resolution of 760\,fA.  The EIS module frequency range is 7\,MHz (3\%, 3$^{\circ}$) down to 10\,$\mu$Hz; 3\,MHz (1\%, 1$^{\circ}$). 

\section{Results and discussion}

Experimental analysis of electrochemical capacitors is usually carried out based on their cyclic voltammetry response, i.e. current or accumulated charge vs. triangular voltage input.  
In Fig.\;\ref{qvt}(a) we show the accumulated charge    on the device (computed from the time-integral of the current, also shown in the figure) in response to a single linear voltage ramp from 0 to 2.5\,V at the voltage rate of 0.1\,V\,s$^{-1}$. The data points are downsampled for computational purposes, and limited to the ones collected every 0.1\,s only.  Here we analyze one half of a cycle (charging sequence), but the same can be done for the subsequent discharging ramp. From a first glance we can see that the charge is not perfectly    proportional to the applied voltage \cite{QV}, indicating that the capacitance cannot be treated simply as a constant, but rather implicitly, as a time-dependent function.

The steps for extracting the capacitance function from the charge and voltage vectors are as follows. First, in Fig.\;\ref{qvt}(b), we show for visualization purposes the sparsity pattern of the matrix $\mathbf{v}$ (Eq.\;\ref{matrixv}, of size 248$\times$248) constructed from the voltage time series data. Nonzero-valued elements are colored in blue while zero-valued elements are displayed in white. As mentioned above this is the lower triangular matrix of the Toeplitz matrix whose columns are shifted versions of the time series voltage signal.

Second, we analyze the voltage matrix by SVD method \cite{hansen2007regularization}. Plots of singular values $\sigma_i$ and the SVD-components $|u_i^{\text{T}}\mathbf{q}|$ and $|u_i^{\text{T}}\mathbf{q}/\sigma_i|$ \cite{hansen1990discrete} of the matrix $\mathbf{v}$  are shown in Fig.\;\ref{qvt}(c). The singular values decay gradually and smoothly from $\sigma_1=\text{177.89}$ to $\sigma_{248}=\text{0.0063}$ with $\text{cond}(\mathbf{v})=2.8246\times 10^4$. The absolute values of the SVD coefficients of the right-hand side, i.e. $|u_i^{\text{T}}\mathbf{q}|$, show an  initial decay  for  $1\leqslant i\leqslant 20$  faster than that of $\sigma_i$ (gradients are plotted in Fig.\;\ref{qvt}(d)) indicating that our problem satisfies the discrete Picard condition in this relatively narrow region \cite{hansen1990discrete, hansen2007regularization}. 
However, if $\sigma_i$ decayed faster than $|u_i^{\text{T}}\mathbf{q}|$, then with an even small perturbation in the measurements one should expect a highly fluctuating solution dominated by the noise.  
We  also note that for the same region of $i\leqslant 20$  the terms $|u_i^{\text{T}}\mathbf{q}/\sigma_i|$ are decaying, but at a slower rate than $|u_i^{\text{T}}\mathbf{q}|$. 
For $i>20$, $|u_i^{\text{T}}\mathbf{q}|$ show a highly fluctuating behavior but the overall trend is somehow leveling off at close to the noise level of the measurements, whereas the terms $|u_i^{\text{T}}\mathbf{q}/\sigma_i|$ show first a slight increase followed by fluctuations at around 10$^{-3}$--10$^{-2}$ in value.

Plots of the Tikhnov-regularized solutions $\mathbf{\tilde{c}}_{\lambda}$ for different values of $\lambda$ (Eq.\;\ref{eq:clambda}) logarithmically-spaced from the lowest to the highest singular value (i.e. from 0.0063 to 177.89) are shown in Fig.\;\ref{qvt}(e). Plots of the associated filter factor terms, $f_i= {\sigma_i^2}/{(\sigma_i^2+\lambda^2)}$ in Eq.\;\ref{eq:clambda}, are shown in Fig.\;\ref{qvt}(f);  for $\sigma_i\gg \lambda$, $f_{i}\to 1$ and for $\sigma_i\ll \lambda$, $f_{i}\to \sigma_i^2/\lambda^2$. 
The overall trend of the capacitance function follows a power-law decay with time that is clearly observed for    small values of $\lambda$, whereas for  large values of $\lambda$, $\mathbf{\tilde{c}}_{\lambda}$ is greatly damped, and tends to zero. 
Finally, in  Fig.\;\ref{qvt}(g) we show   the regularized solution for the particular value of $\lambda=0.0403$. This value of $\lambda$ is obtained via the L-curve method (not shown here)  \cite{hansen2007regularization}, and seems to  provide an acceptable balance between perturbation and regularization errors in the solution, knowing that other methods such as the generalized cross-validation criterion may provide different results  \cite{jacquelin2003force}.

\begin{figure}[!t]
\begin{center}
\includegraphics[width=2.2in]{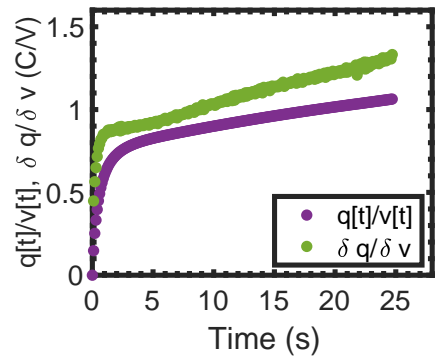} 
\caption{Plot of capacitance of the Samxon EDLC  computed from the ratio of time-domain (i) charge by voltage and (ii) differential charge by differential voltage}
\label{qvct}
\end{center}
\end{figure}

\begin{figure}[b]
\begin{center}
\includegraphics[width=2.2in]{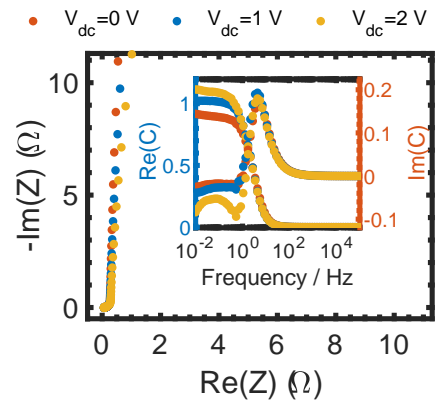} 
\caption{Plot of spectral impedance data of the Samxon EDLC device in the Nyquist form of real vs. imaginary parts of impedance for different voltage dc biases; inset shows plots of real  and imaginary part of the frequency-domain capacitance function given by Eq.\;\ref{eq21}}
\label{eis}
\end{center}
\end{figure}

\begin{figure}[!t]
\begin{center}
\includegraphics[width=2.2in]{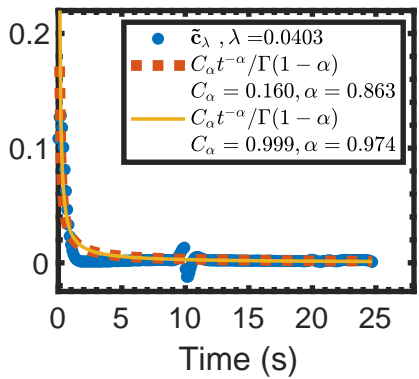} 
\caption{Plot of time-domain capacitance functions obtained by (i) Tikhnov regularization of deconvolution problem of time-domain charge-voltage data, and from inverse Laplace transform of frequency-domain fractional-order impedance model}
\label{ct}
\end{center}
\end{figure}

In comparison with the capacitance as commonly computed from the simple point-by-point division of charge by voltage or the division of differential change in charge by change in voltage,  as shown in Fig.\;\ref{qvct}, it is clear that there are discrepancies between the two resulting capacitance functions \cite{allagui2021inverse, ieeeted, acs2}. 
To further look into this, we collected the spectral impedance of the device at different dc voltage biases from 0 to 2.5\,V  with stepping sine excitations of   14.14\,mV amplitude over the frequency range 10\,mHz to 100\,kHz. Some of the data are plotted in Fig.\;\ref{eis} in terms of real vs. imaginary parts of impedance, showing little effect of applied dc biases for this device. The low-frequency capacitive branch from \emph{ca.} 1\,Hz down to 10\,mHz can  satisfactorily be fitted with a constant phase impedance defined as:
\begin{equation}
Z(s) = \frac{V(s)}{I(s)} = \frac{V(s)}{s\,Q(s)} = \frac{1}{s^{\alpha} C_{\alpha}}
\label{eq20}
\end{equation}
 where $\alpha$ is a real coefficient that takes values between 0 and 1, and $s= j 2\pi f$. For example, for the case of 2.0\,V dc bias, the fitting parameters (using complex nonlinear least-squares data fitting) were found to be $C_{\alpha}=0.999$\,F\,s$^{\alpha-1}$ and $\alpha=0.974$ with $\chi^2=0.068$. The effect of dc bias on these model parameters is also relatively small.  
From Eq.\;\ref{eq20} a frequency-dependent, not constant,  capacitance function is computed from the ratio of charge by voltage as:
\begin{equation}
C(s) = \frac{Q(s)}{V(s)} = C_{\alpha} s^{\alpha-1}
\label{eq21}
\end{equation}
given that we were dealing with convolution operations in the time domain. Plots of the real and imaginary parts of  capacitance as a function of frequency for the three different dc biases are shown in the inset of Fig.\;\ref{eis}.  The inverse Laplace transform (defined as 
$(2\pi i)^{-1}\int_{\gamma-j\infty}^{\gamma+j\infty} F(s)e^{st} ds$) of $C(s)$ giving the corresponding time-domain capacitance is a decaying power law function, i.e.:  
\begin{equation}
c(t) = \frac{C_{\alpha} t^{-a}}{\Gamma(1-\alpha)}
\label{eqCt}
\end{equation}
which is in accordance with the Tikhonov-regularized solution depicted in Fig.\;\ref{qvt}(e) for example. 
In  Fig.\;\ref{ct}, we replot  
$\mathbf{\tilde{c}}_{\lambda}$ with $\lambda=0.0403$
  along with the   fitting model function $c(t)$ (Eq.\;\ref{eqCt}) with the best fitting parameters 
 ($C_{\alpha}=0.160$\,F\,s$^{\alpha-1}$, $\alpha=0.863$, using nonlinear  least-squares data fitting), and with  
 ($C_{\alpha}=0.999$\,F\,s$^{\alpha-1}$, $\alpha=0.974$) from impedance spectroscopy analysis, noting that the model parameters  extracted from time-domain and frequency-domain   data modeling should not be necessarily  the same. Besides the effect of the regularization procedure, this explains the quantitative difference between the time-domain capacitance functions presented in Fig.\;\ref{ct}.

Thus, the discrepancies between the time-domain capacitances computed by division of charge by voltage as shown in Fig.\;\ref{qvct} vs. deconvolution as presented in this work (Fig.\;\ref{ct}) is due to confusion between definitions \cite{acs2}. In this study, we assumed that the recommended formula for computing the capacitance in the frequency domain is $Q(s)=C(s) V(s)$ given that it is a direct extension of the very definition of impedance. From there, the time-domain relationship between charge and voltage should be the convolution $q(t) = (c \circledast v) (t)$, and not $q(t) = c (t) v (t)$.  

Finally, we note that the same analysis shown above in Fig.\;\ref{qvt} for the case of charging  linear voltage ramp has been repeated for the same device with a constant charging voltage. We obtained comparable results that are not shown here to avoid repetition and redundancy. 

\section{Conclusion}

The capacitance function of a capacitive energy storage device cannot be directly measured, but can be estimated from input and output signals. In this study we showed how to treat the ill-posed convolution problem of capacitance-by-voltage giving the accumulated charge on a non-ideal capacitive device that can be corrupted with noise. We used the Tikhonov regularization method, and verified the concordance between the regularized solution obtained from time-domain data and the one obtained from frequency-to-time transformation of the frequency-domain capacitance. The latter is derived from the definition of impedance being the ratio of frequency-domain voltage by current.
   
%\section*{Acknowledgement}
%
%This work was supported by the University of Sharjah (Project \# 1602040634-P). 

\section*{References}

\bibliographystyle{naturemag}
%\bibliography{bib}


%\section*{Author contributions statement}

%Conceived and designed the experiments
% 		Performed the experiments
% 		Analyzed the data
% 		Contributed materials/analysis tools
% 		Wrote the paper

%\section*{Additional information}
%\subsection*{Competing interests}
%The authors declare no competing interests.

%\section*{Data Availability}
%
%The data that support the findings of this study are available from the corresponding author upon request.

 \end{document}